# LOCOFY Large Design Models - Design to code conversion solution

**Sohaib Muhammad[*1], Ashwati Vipin[*#2], Karan Shetti[1], Honey Mittal[1]**

Affiliations:

[1]Locofy.ai | Locofy Pte. Ltd.

[2]Lee Kong Chian School of Medicine, Nanyang Technological University, Singapore

*Joint first authors.

[#] Corresponding Author

Correspondence email: ashwati.vipin@ntu.edu.sg

# ABSTRACT

Despite rapid advances in Large Language Models and Multimodal Large Language Models (LLMs), numerous challenges related to interpretability, scalability, resource requirements and repeatability remain, related to their application in the design-to-code space. To address this, we introduce the Large Design Models (LDMs) paradigm specifically trained on designs and webpages to enable seamless conversion from design-to-code. We have developed a training and inference pipeline by incorporating data engineering and appropriate model architecture modification. The training pipeline consists of the following: 1)Design Optimiser– developed using a proprietary ground truth dataset and addresses sub-optimal designs; 2)Tagging and feature detection– using pre-trained and fine-tuned models, this enables the accurate detection and classification of UI elements; and 3)Auto Components– extracts repeated UI structures into reusable components to enable creation of modular code, thus reducing redundancy while enhancing code reusability. In this manner, each model addresses distinct but key issues for design-to-code conversion. Separately, our inference pipeline processes real-world designs to produce precise and interpretable instructions for code generation and ensures reliability. Additionally, our models illustrated exceptional end-to-end design-to-code conversion accuracy using a novel preview match score metric. Comparative experiments indicated superior performance of LDMs against LLMs on accuracy of node positioning, responsiveness and reproducibility. Moreover, our custom-trained tagging and feature detection model demonstrated high precision and consistency in identifying UI elements across a wide sample of test designs. Thus, our proposed LDMs are a reliable and superior solution to understanding designs that subsequently enable the generation of efficient and reliable production-ready code.

# INTRODUCTION

In the current vast digital landscape, the battle for user attention is fierce, and time to market, as well as the quality of a website's user interface (UI) and user experience (UX), play a crucial role in its success. However, the challenges in creating engaging and functional websites are substantial.

Studies indicate that 42% of consumers leave a website when encountering usability problems or poor functionality.[1] On the other hand, well-executed UI designs can increase website conversion rates by 200%, while optimized UX design can boost these rates by 400%. However, many businesses need help bridging the gap between great designs and their final products. To combat these challenges, tools that enable bridging the gap between design and end product are needed.

Indeed, there has been rapid growth in the volume of text-related data coupled with the enormous advance in Large Language Models (LLMs) [Open AI ChatGPT, GPT4] and Multi-modal Large Language Models (LMMs) [OpenAI. GPT-4V(ision) System Card.]. While a large array of problems have been addressed using these models, many challenges remain related to their use in the design-to-code space. These include, but not limited to, scalability, resource requirements as well as interpretability.[2,3,4]

In particular, the effective conversion of designs into code by Large Language Models (LLMs) poses challenges, as they are primarily trained on textual data, making it difficult for them to understand the visual elements of design, leading to less precise code generation when translating complex visual designs into functional code.[5]

On the other hand, LMMs such as GPT-4V and Gemini Vision Pro have shown notable progress in addressing the limitations of traditional LLMs in design-to-code conversion. These models integrate visual and textual understanding, enabling them to process and generate content across different media types. This multimodal capability empowers LMMs to effectively interpret visual designs and generate corresponding code, outperforming text-only models in such tasks.[6,7]

Despite these advancements, LLMs and LMMs still face challenges in achieving perfect conversion from design to code. While they can generate designs and associated code, they are unable to take existing designs and convert them into accurate code. They often encounter issues with ensuring the generated code is optimized, maintainable, and responsive across different devices and screen sizes. Front-end code involves not only translating visual elements but also ensuring proper layout, responsiveness, and user interaction, which requires a deep understanding of how UI elements are designed and structured and not just a deep understanding of CSS, HTML and Javascript.. In this manner, LLMs/LMMs may generate syntactically correct code that however eventually fails to meet design expectations regarding visual fidelity and functionality.[6,7]

With the goal of seamlessly converting designs into correct and usable code, we introduce the Large Design Model (LDM) paradigm—purpose-built to understand and translate design files into production-ready front-end code. Unlike LLMs trained on text or images, LDMs are trained on actual design files and web pages, giving them a native understanding of structure, hierarchy, and interaction patterns. Our models directly process design metadata, layer structures, and visual context to infer layout behaviour, tag UI elements, and identify reusable components. This multimodal processing allows LDMs to grasp the functional intent behind graphical elements and generate consistent, modular, and maintainable code. Trained on a corpus of over one million web pages — and continuing to scale — this model builds a comprehensive understanding of design. Identifying interactive UI elements and optimizing the design structure for optimal code generation serves as the foundational model's primary application. We also train specialized models for downstream tasks such as UI element tagging, component detection, feature recognition and responsiveness. The contributions of this framework include a novel model architecture and optimization strategies for LDMs, enabling a seamless transition from design to code.

## LARGE DESIGN MODELS (LDMs)

LDMs are specific to design paradigms through their exclusive training on websites and designs in order to attain model sensitivity and specificity. Thus, in the larger landscape of LLMs and similar frameworks, LDMs are uniquely positioned to deal with all design to code needs because of their specialised training. In order to ascertain this we have conducted a series of experiments using existing frameworks as well as the Locofy LDM framework. Our key innovation comes from how we use data that is suitable for the design space. Thus our approach uses a combination of data engineering and modifying the right model architecture to enable the LDM framework (Figure 1).

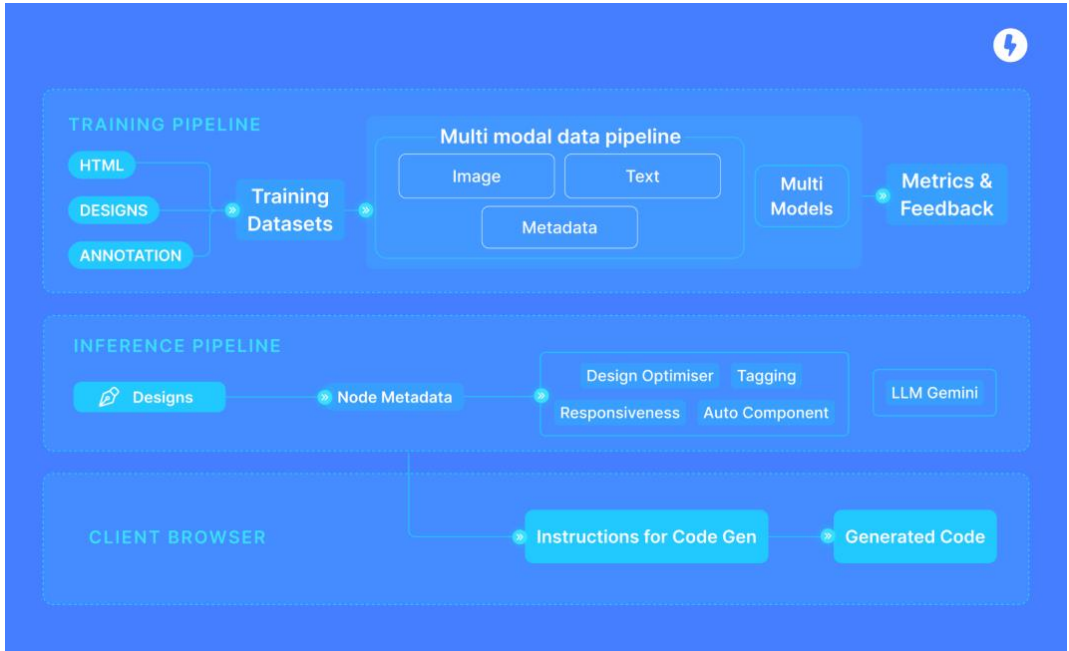

**Figure 1: Overall architecture comprising the training pipeline and inference pipeline.**

*Training Pipeline*

The training pipeline for our Large Design Model (LDM) is built around a multimodal approach that utilizes diverse, curated datasets, which include design metadata, text and image information(Figure 2). These datasets, which are labeled and curated in-house, form the foundation for training the LDM. By incorporating metadata extracted directly from design layers, the model learns rich, multi-view representations that enhance the accurate recognition and classification of design components.

This embedding-driven framework allows the model to generalize across a wide range of UI elements and serves as a basis for expanding classification capabilities to more complex and custom interface patterns, including carousels, image galleries, and accordions.

The overall system consists of four core models: Design Optimizer, Tagging, Auto Components, and Responsiveness. Each model addresses different aspects of the design-to-code transformation process. Depending on the specific task, we utilize a combination of Transformer-based architectures, deep learning techniques, and traditional machine learning models to enhance performance across various use cases.

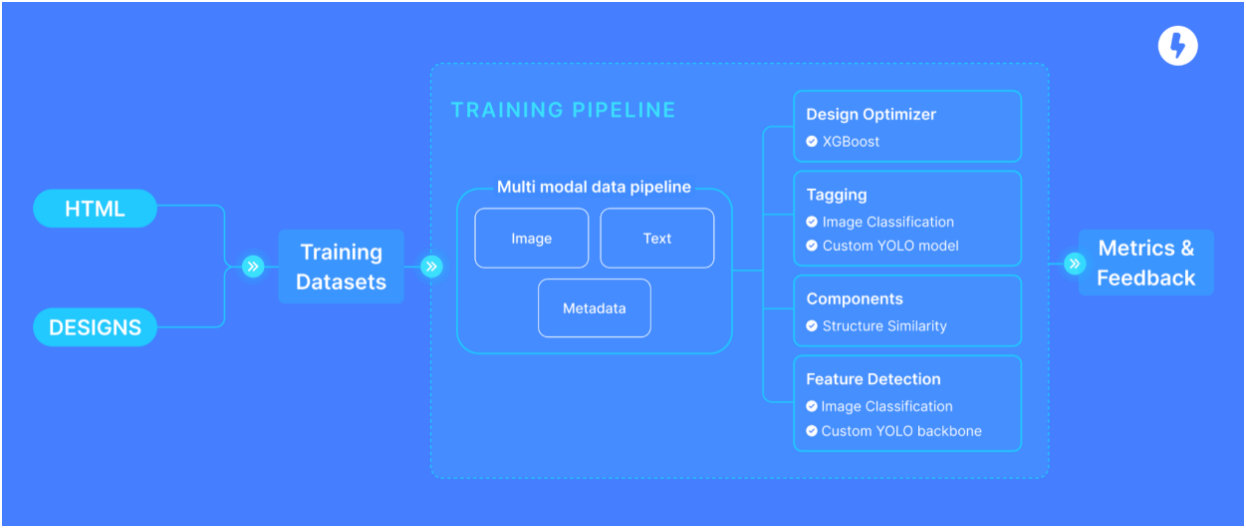

Figure 2: LDM training pipeline

*Inference Pipeline*

The LDM is part of a deterministic inference pipeline(Figure 3) that processes real-world design files to create structured, intermediate instructions for code generation (see Figure 3). Unlike traditional generative approaches, our system does not directly synthesize code from model outputs. Instead, it produces a set of precise and interpretable instructions that are utilized by our proprietary code generation engine. This separation ensures that the same design consistently produces the same code output, a critical requirement for reliability and version control in professional workflows.

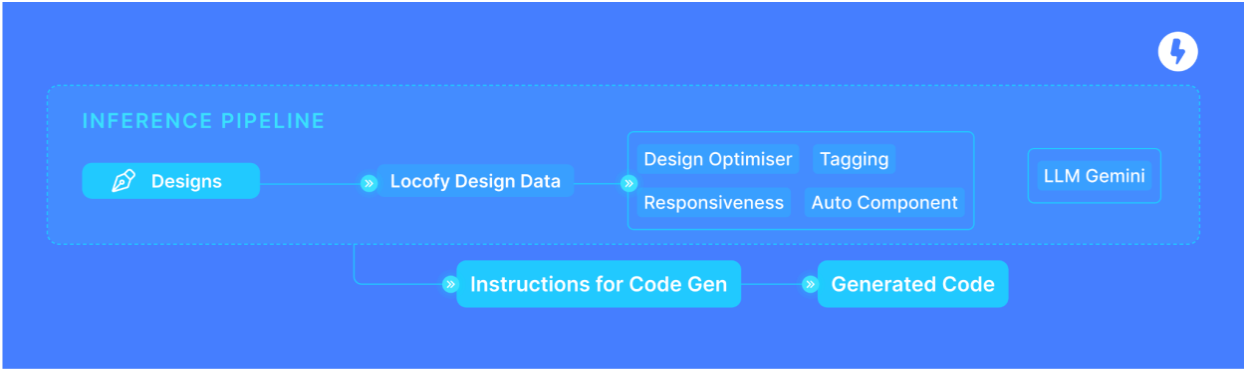

Figure 3: LDM inference pipeline

In the following sections, we detail the components of the LDM model architecture.

*Design Optimiser*

Problem Statement: The Design Optimizer module addresses the common problem of unstructured and suboptimal design files, which represent a significant portion of real-world design data (see Figure 4). This process starts with a step that groups layers (see Figure 5), allowing for the identification of relational hierarchies between elements and thus enables the use of higher-level properties, such as Auto Layout (Figure 6), which helps produce cleaner, more maintainable, and responsive code.

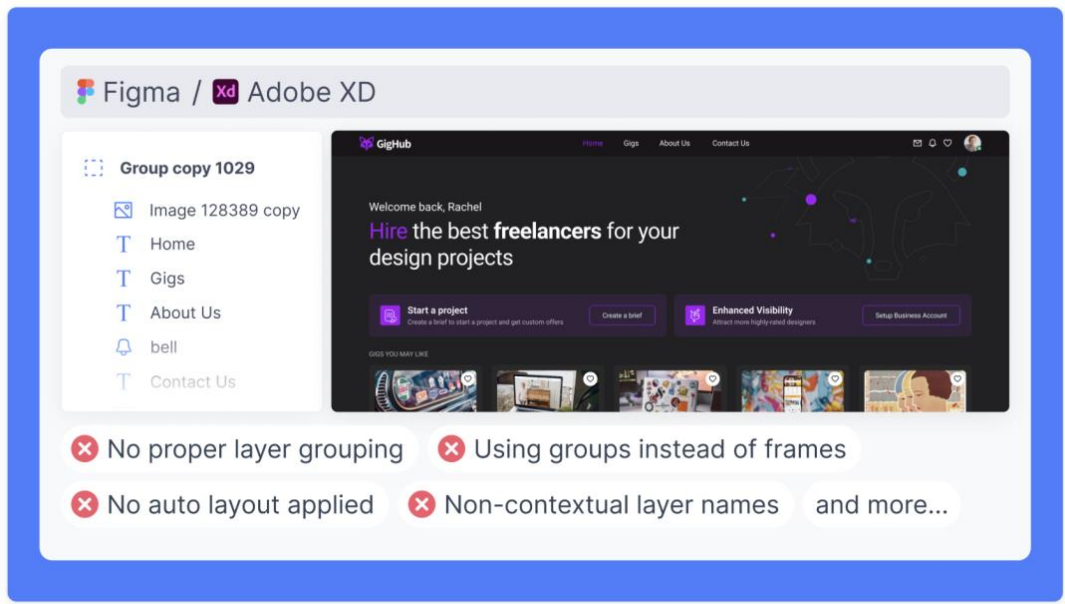

**Figure 4: Features of a sub-optimal design**

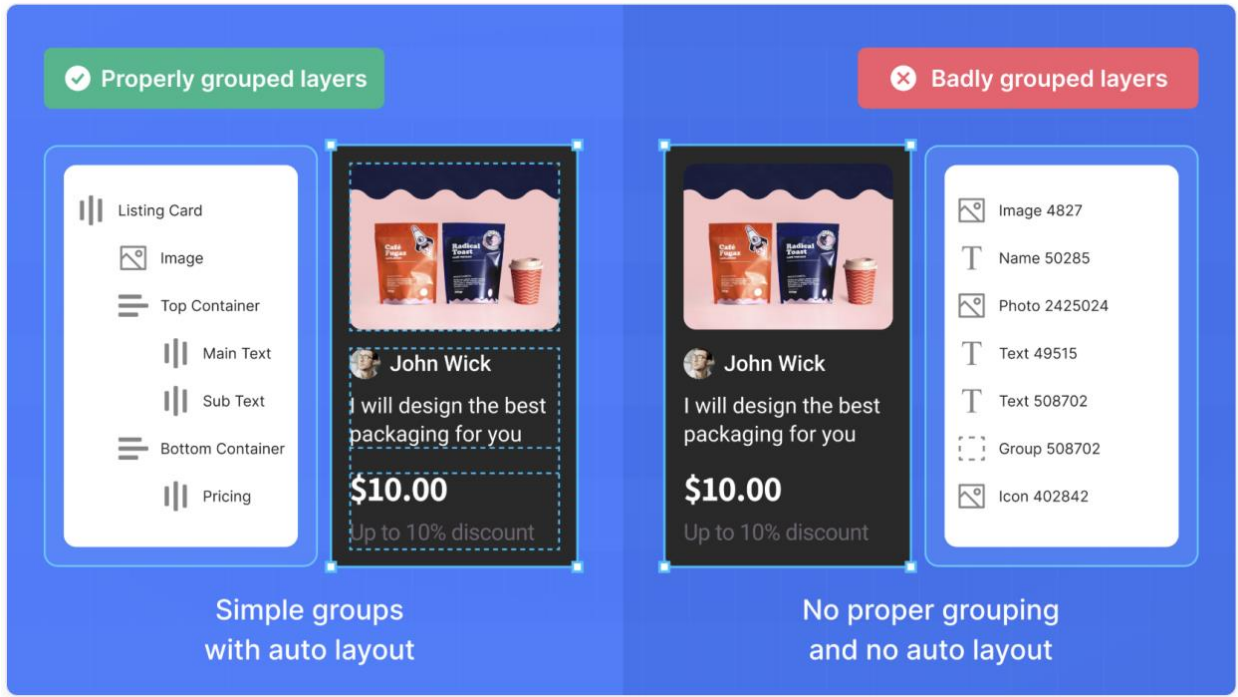

Figure 5: Grouping of layers

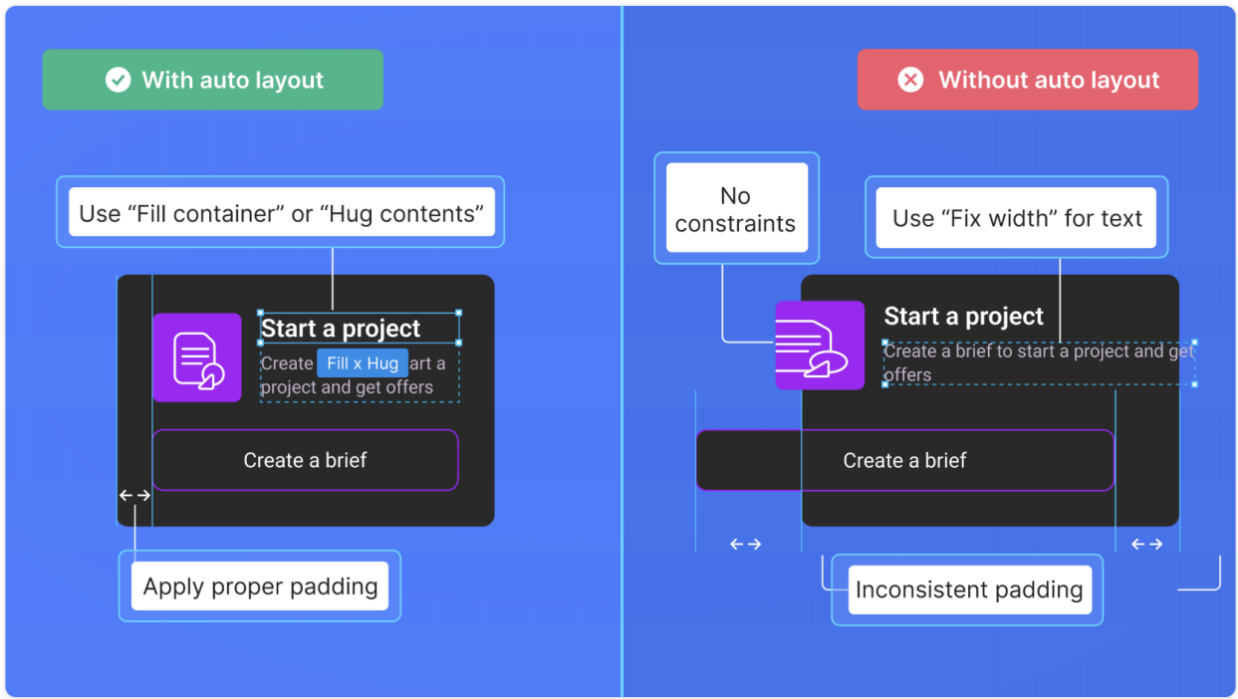

Figure 6: Auto layout enables designs to stay interactive at any form factor.

Approach: To develop the Design Optimizer, we created a proprietary ground truth dataset. This involved collecting poorly structured designs and then working with expert designers to annotate the best-practice transformations. These annotations provided the basis for supervised training using an XGBoost-based framework, utilizing these expert annotations to learn how to map unoptimized design states to optimized ones. This optimization framework supports responsiveness across multiple screen sizes and ensures that the final designs are semantically grouped and structurally aligned, enhancing interactivity and improving the quality of the resulting code.

*Tagging and Feature Detection*

Problem Statement: A fundamental challenge in design-to-code systems is that design files often represent all visual elements—such as buttons, inputs, and headers (Figure 7)—as indistinguishable geometric shapes (e.g., rectangles, vectors). This abstraction obscures the underlying semantic and interactive intent of UI components, making it difficult to generate code that aligns with the expected functionality across different front-end libraries and frameworks. As a result, accurate detection and classification of these interactive elements are critical for producing high-quality, context-aware code.

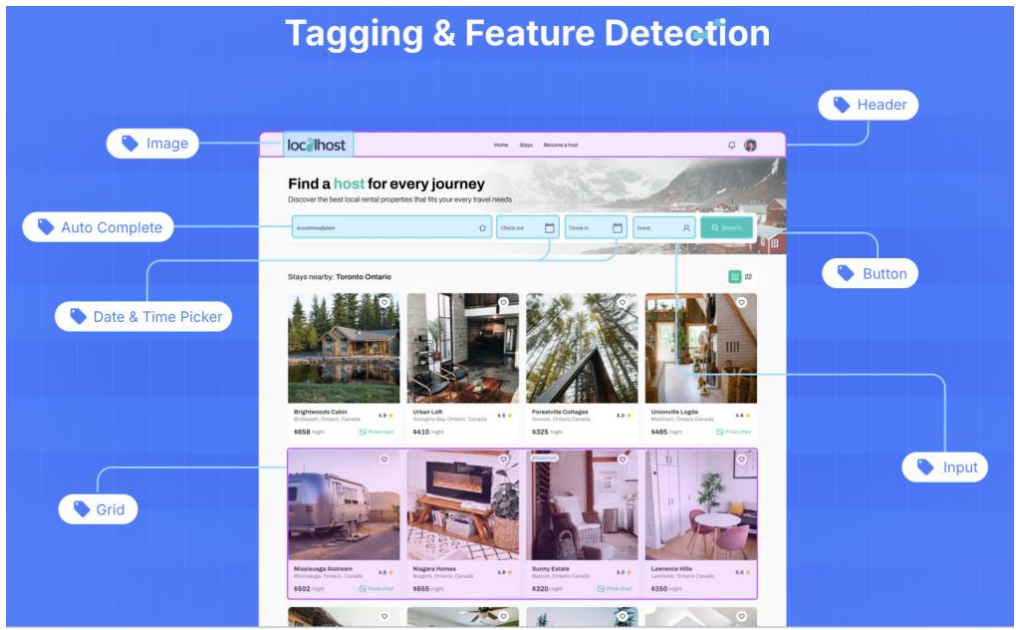

**Figure 7: Tagging and Feature Detection**

Approach: We present a Tagging and Feature Detection model based on a YOLO backbone to address this issue. This model can identify individual UI features—such as buttons, input fields, and icons—and combine them into more complex structures like headers, cards, or navigation groups (see Figure 7). Importantly, in addition to tagging elements at a basic level, the model also reorganizes and groups components based on visual hierarchy and spatial relationships within the design. This structural interpretation ensures that detected elements are accurately identified and organized to reflect their proper functional grouping.

Jasmine is our custom pre-training strategy designed for the YOLO backbone. Unlike YOLO, which is trained on generic data, Jasmine is pre-trained on a large-scale, UI-specific dataset curated from over 1 million web and design nodes. This dataset has been cleaned and organized to highlight common UI elements such as images, frames, and rectangles, allowing the model to better internalize the unique visual patterns found in digital interfaces. After pretraining, the model undergoes fine-tuning on annotated design datasets to accurately classify the specific UI tags needed for downstream design-to-code tasks.

Within the broader LDM framework, this enhanced representation is utilized to generate consistent and deterministic code instructions. Understanding element semantics and their structural relationships is crucial for producing clean, modular, and interactive code—especially when focusing on component-based UI libraries and ensuring long-term maintainability.

*Auto Components*

Problem Statement: Another significant challenge in large-scale design-to-code systems is the redundancy and inconsistency that results from repeated UI patterns across multiple screens (Figure 8). In design files, recurring elements—such as input fields, cards, or product tiles—are often duplicated instead of being abstracted into reusable components. This lack of modularity leads to bloated, repetitive code, which hinders the maintainability and scalability of the generated frontend. Furthermore, without a systematic approach to identify and abstract these repeated elements, ensuring consistent behavior and styling across instances becomes difficult, as does the efficient integration of backend logic.

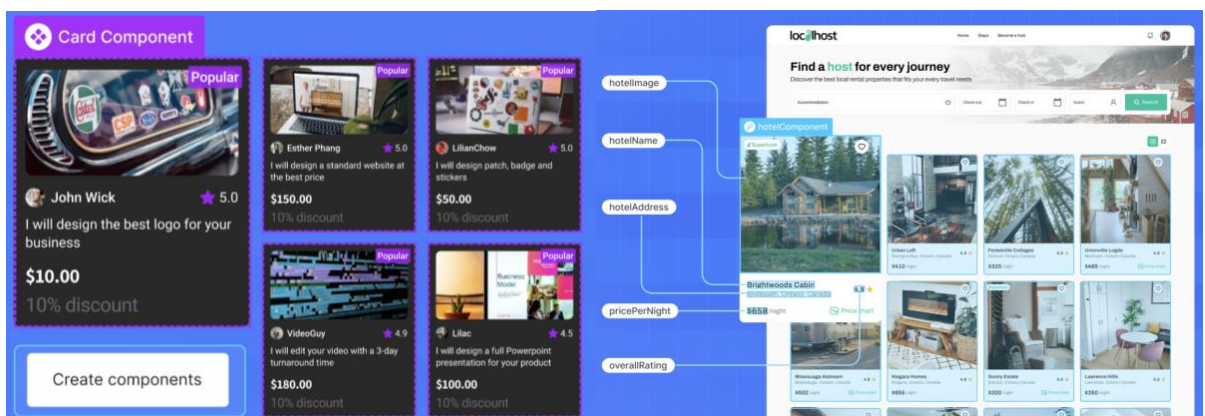

**Figure 8: Auto components for reusable elements: Components help maintain consistency and reduce redundancy by allowing users to reuse UI elements across multiple screens.**

Approach: To tackle this issue, we present the Auto Component model, which is specifically designed to identify and extract repeated UI structures into reusable components throughout the design space. The module detects visually and semantically similar groups of elements—both within individual screens and across different screens—and conveys this grouping to the code generator as component-level abstractions.

This approach enables the creation of modular code, allowing a single component definition to be reused multiple times. This practice reduces redundancy and enhances both the readability and reusability of the code. By following component-based frontend principles, the system ensures that code related to repeated elements utilizes shared structures. These structures can be dynamically linked to data and backend services in a consistent manner.

Furthermore, we automatically identify relevant properties (props) for each component and fills them in based on structural patterns and contextual cues from the design. This greatly simplifies the integration of components with backend logic and state management systems, allowing developers to easily connect dynamic data or user interactions without requiring extensive manual setup. This abstraction not only promotes design consistency and concise code but also improves development efficiency by enabling seamless, production-ready integration of UI components with backend endpoints.

# COMPARATIVE EXPERIMENTS AND DISCUSSION

To examine model performance, we utilised several evaluation metrics as detailed below.

*End-to-end design to code conversion accuracy*

To quantitatively evaluate the performance of our LDM, we introduce a novel metric—Preview Match Score—which measures the visual fidelity between the original design and the model-generated output. During the model development phase, comparisons were performed at a node-level granularity, assessing discrepancies in width, height, and absolute (x, y) coordinates between the source design and the rendered output.

To compute the **Preview Match Score** for a given design screen, we evaluate each node $i \in \{1, 2, ..., N\}$ in the original design and its corresponding rendered output. For each node, we compare its position and dimensions:

- Original: $(x_i, y_i, w_i, h_i)$
- Rendered: $(\hat{x}_i, \hat{y}_i, \hat{w}_i, \hat{h}_i)$

A node is considered a **match** if the **absolute percentage error** for **all** four attributes is within a threshold of 3%:

$$\left|\frac{x_i - \hat{x}_i}{x_i}\right| \leq 0.03, \quad \left|\frac{y_i - \hat{y}_i}{y_i}\right| \leq 0.03, \quad \left|\frac{w_i - \hat{w}_i}{w_i}\right| \leq 0.03, \quad \left|\frac{h_i - \hat{h}_i}{h_i}\right| \leq 0.03$$

Let $M$ be the number of nodes that satisfy this condition. Then the **Preview Match Score** for the screen is defined as:

$$\text{PMS} = \frac{M}{N} \times 100$$

Where:

- $N$ is the total number of nodes in the design,
- $M$ is the number of nodes correctly matched within the 3% tolerance.

This score represents the percentage of nodes in a screen whose layout and dimensions closely match the intended design, providing a robust measure of visual fidelity in rendered output.

The objective of this metric is to minimize visual mismatch on a per-screen basis, with a target of achieving a near-perfect match. Across a test set of 1,000 real-world designs collected from the community, our models consistently attained high fidelity, with 89.6% of screens achieving a Preview Match Score exceeding 95%. This implies that over 95% of nodes across these screens perfectly align in spatial and dimensional attributes with their original design counterparts. The distribution of scores across all tested screens is shown in Figure 9.

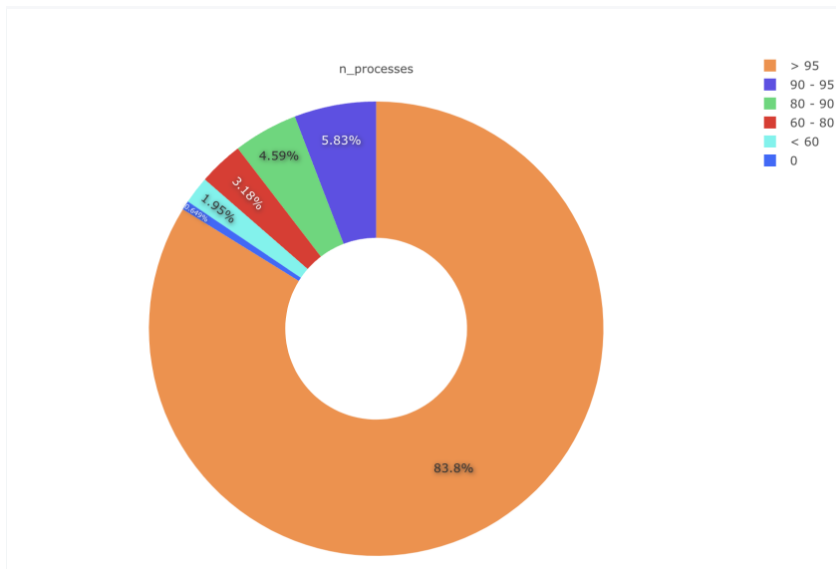

**Figure 9: Distribution of preview match scores across different screens**

Furthermore, in scenarios where precise visual alignment is critical, users can leverage Locofy Edit Mode to achieve 100% preview match, thereby enabling pixel-perfect code generation. This metric also served as a benchmark in our comparative analysis against general-purpose LLM-based approaches. In these evaluations, alternative LLMs consistently underperformed, exhibiting significantly lower preview match scores, highlighting the superiority of LDMs in preserving spatial accuracy and design fidelity.

*Design Optimizer*

To evaluate the performance of our Design Optimizer, we conducted a series of comparative experiments using three different modeling strategies: (1) general-purpose LLMs, (2) fine-tuned LLMs, and (3) our proposed LDM. These experiments aimed to assess the accuracy of node positioning and the reconstruction of hierarchy in design files of increasing complexity.

The evaluation was conducted on a carefully curated benchmark of unoptimized designs and their optimized results featuring various levels of nesting and structural depth (see Figure 10). The comparative baselines included Meta-LLaMA, Google T5, Gemini 1.5 Pro, and ChatGPT-4o. As the number of nested layers increased from 2 to 3, many large language models (LLMs) struggled to produce coherent outputs, often yielding structurally invalid or incomplete hierarchies. This deterioration in performance intensified with deeper nesting, primarily due to token limitations, mis grouping of elements, and a failure to maintain layout constraints.

| Model Name | Number of Nested layer | 2 | 3 | 4 | 5 | 5 | 6 | 6 | 7 | 7 |
|---|---|---|---|---|---|---|---|---|---|---|
| | | 2780 | 6385 | 6344 | 6610 | 13364 | 10156 | 17432 | 13111 | 12454 |
| meta-llama/Llama-2-7b-hf | | Returned | No result | No result | No result | No result | No result | No result | No result | No result |
| meta-llama/Llama-3.1-8B | | Returned results | No result | No result | No result | No result | No result | No result | No result | No result |
| google-t5/t5-11b | | Returned results | No result | No result | No result | No result | No result | No result | No result | No result |
| Gemini-1.5-pro | | Returned results | Returned results | Returned results | Returned results | Returned results | Returned results | No result | No result | No result |
| chatgpt-4o | | Returned results | Returned results | Returned results | Returned results | No result | No result | No result | No result | No result |

**Figure 10: LLM vs LDM testing results**

In contrast, our LDM demonstrated strong performance across all complexity levels, consistently generating structurally accurate and hierarchically aligned outputs. We also assessed next-token prediction accuracy, which measures a model's ability to predict the correct continuation in a design-to-code sequence. The LDM outperformed the others on this metric as well, particularly with deeply nested designs.

*LLM model finetuning*

To further assess the effectiveness and scalability of large language models (LLMs) in this context, we conducted targeted fine-tuning experiments. Our goal was to evaluate both cost efficiency, measured as cost per node, and token conversion performance. The first experiment utilized a text-only setup, while the second included images, text, and associated metadata. Although the multimodal configuration showed slightly better performance, it incurred significantly higher costs per node, rendering it impractical for large-scale deployment (Figure 11).

The setup and configurations for these experiments are summarized in Table 1. Importantly, by applying parameter-efficient fine-tuning (PEFT) to the Gemini 1.5 Pro model, we found that a text-only dataset with 6,000 samples outperformed a multimodal dataset with 1,000 samples. This indicates that large-scale textual supervision alone can be both cost-effective and efficient in this domain.

Table 1 : LLM finetuning setup

|  | Gemini 1.5 Pro - Text | Gemini 1.5 Pro - Image + Text |
|---|---|---|
| Base model | gemini-1.5-pro-002 | gemini-1.5-pro-002 |
| Tuning Method | Supervised | Supervised |
| No of epochs | 10 | 5 |
| Adaptor size | 4 | 4 |
| Dataset | 6000 input, output pairs | 1000 input, output pairs |

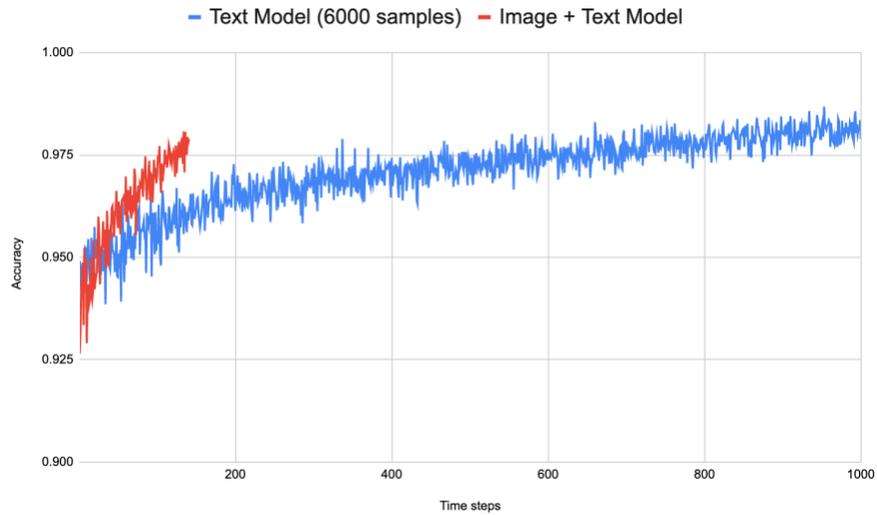

Figure 11: Next token prediction accuracy

Competing LLM-based models frequently encountered issues such as:

- Lack of component structure
- Poor responsiveness
- High mismatch scores during previews
- Low consistency and reproducibility in the generated output code

In contrast, Locofy's LDM achieved superior results in just a few seconds, producing repeatable, high-quality, and component-consistent code as measured by our preview match score.

*Qualitative results*

The following shows the qualitative results comparing the best performing LLMs and our LDM

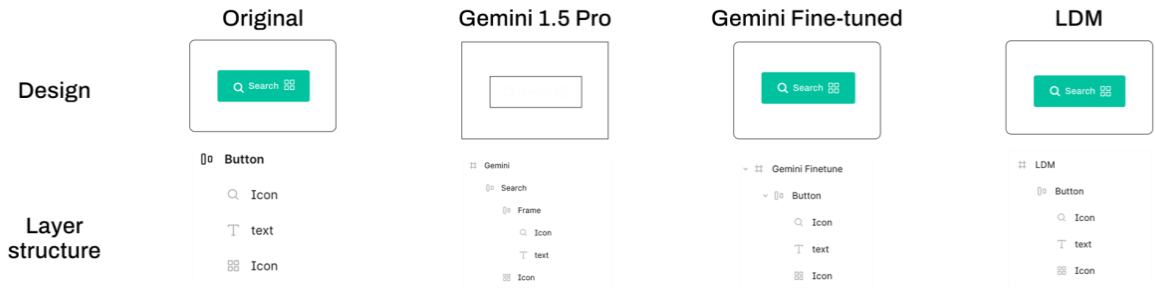

Example 1 : 2 layer nested tree

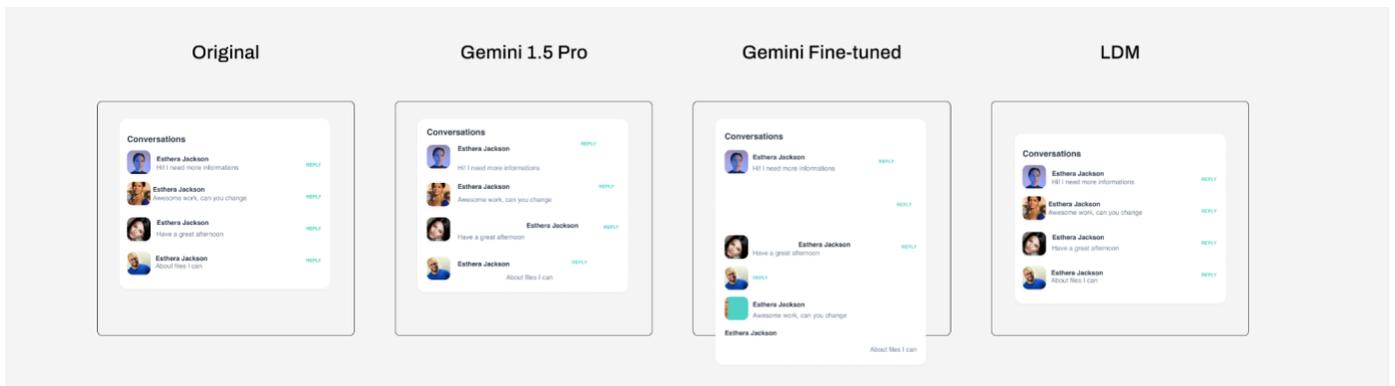

Example 2 : 5 layer nested tree

*Tagging and feature detection performance*

In the next phase of our evaluation, we conducted a series of experiments to assess the accuracy and evolution of our tagging and feature detection models. We specifically compared baseline YOLO models with our custom-trained Jasmine model, using standard classification metrics—precision, recall, and F1-score—across both small and large tag categories (see Table 1).

Initial production results from Resnet models showed relatively low macro-average F1 scores: 32.78% for small tags and 14.33% for large tags. However, significant improvements were observed when we switched to the YOLO-based model, with F1-scores increasing to 76.09% for small tags and 51.39% for large tags. The introduction of Jasmine, which was pretrained on UI-specific data and fine-tuned on curated annotations, resulted in even greater performance gains, achieving F1-scores of 85.79% for small tags and 71.40% for large tags.

We also tested an augmented version of Jasmine that incorporated text-based features, which maintained comparably high performance. Nevertheless, the best results were achieved with the post-processed Jasmine model, which reached macro-average F1-scores of 86.07% for small tags and 77.22% for large tags (Table 2). This model was selected as the final version for our tagging and feature detection pipeline.

Across a wide range of 5,000 test designs, the chosen model demonstrated high precision, strong generalization, and consistent performance, reliably identifying UI elements with production-grade accuracy.

Table 2 : Tagging and feature detection performance

|  | Tag | YOLO F1 Score | Jasmine F1 Score |
|---|---|---|---|
| Small Tags | Date time picker | 61.58% | 63.99% |
| Small Tags | button | 81.16% | 96.95% |
| Small Tags | select | 81.90% | 91.24% |
| Small Tags | input | 88.91% | 92.34% |
| Small Tags | checkbox | 73.34% | 94.77% |
| Small Tags | radio | 70.41% | 94.27% |
| Small Tags | textarea | 83.97% | 79.68% |
| Small Tags | dropdown | 68.62% | 70.34% |
| Small Tags | switch | 74.91% | 91.01% |
| Big tags | Audio_player | 23.51% | 59.24% |
| Big tags | Drawer | 34.82% | 91.60% |
| Big tags | File_upload | 56.55% | 50.17% |
| Big tags | Google_maps | 57.68% | 90.22% |
| Big tags | Grid | 80.79% | 95.67% |
| Big tags | Popups | 68.00% | 89.09% |
| Big tags | Progress | 10.15% | 61.57% |
| Big tags | Slider | 59.88% | 71.00% |
| Big tags | Video | 52.98% | 86.75% |
| Big tags | Quantity_selector | 69.50% | 76.88% |
|  | Macro average small tags | 76.09% | 86.07% |
|  | Macro average big tags | 51.39% | 77.22% |

**SUMMARY**

Thus, our performance analysis illustrated that the design-to-code space requires a tailor-made solution such as our LDMs. We show that LLMs lack in performance and are not designed or optimised to solve design to code conversion. Moreover, we would like to highlight that at the point of inception of our LDMs, LLMs were unable to incorporate design files into their model. Through consistent testing and validation of our backend models we have created an in-house design to code solution which is uniquely positioned to bring designs to life.

**LIMITATIONS AND FUTURE WORK**

Our work does have some limitations. For our LDMs to achieve peak performance, the design structure needs to be optimized such that model constraints can be applied and optimised appropriately. Free-form designs and designs generated by AI do not work well for code generation. Additionally, not all users are likely to follow design best practices which may also lead to sub-optimal code generation and user experience and may need manual optimisation. Our LDMs currently have been trained on hundreds of millions of parameters which may not meet the true definition of "Large models". However, we are constantly training our models using increasingly large datasets to improve performance and this will be a major component of all future model releases. We are also working on experiments to further improve performance of our auto-component and responsiveness models. Once designs have been converted to code, they also present an opportunity to further refine or improve using existing LLM frameworks that are trained extensively on code. We are also planning to extend to all design tools as well as web and app design frameworks in the near future.

# CONCLUSION

In this report, we propose LDMs, a novel multimodal approach for design to code automation. LDMs are specific to design paradigms through their exclusive training on web data and designs. Our dataset, curated from over one million web and design nodes, was cleaned and tailored to emphasize common UI structures critical to designs. In this manner, through a comprehensive training and deterministic inference pipeline, LDMs are uniquely positioned to deal with all design to code needs. Our key innovation comes from the way we use data to convert to embeddings that are suitable for the design space. Using our core models and utilising a combination of transformer-based architectures, deep learning and traditional machine learning, we effectively address the complex aspects of the design-to-code transformation process. Our Preview Match Score metric illustrates that our models consistently achieve high fidelity with 90% of screens achieving a >95% match score. Moreover, results from our comparative experiments lend evidence to the superior performance of our LDMs against common LLM frameworks for design optimisation as well as tagging and feature detection. Overall, our models demonstrate high precision and consistent performance for design-to-code conversion and produce repeatable, interactive and high-quality output code.